# Mathematical Modeling of [18]F-Fluoromisonidazole ([18]F-FMISO) Radiopharmaceutical Transport in Vascularized Solid Tumors

Mohammad Amin Abazari[a], M. Soltani[a,b,c,d,e,*], Faezeh Eydi[a], Arman Rahmim[c,f], Farshad Moradi Kashkooli[a]

[a] Department of Mechanical Engineering, K. N. Toosi University of Technology, Tehran, Iran

[b] Department of Electrical and Computer Engineering, Faculty of Engineering, School of Optometry and Vision Science, Faculty of Science, University of Waterloo, Waterloo N2L 3G, Canada

[c] Department of Integrative Oncology, BC Cancer Research Institute, Vancouver, British Columbia V5Z 1L3, Canada

[d] Centre for Biotechnology and Bioengineering (CBB), University of Waterloo, Waterloo, Ontario N2L 3G, Canada

[e] Cancer Biology Research Center, Cancer Institute of Iran, Tehran University of Medical Sciences, Tehran, Iran

[f] Departments of Radiology and Physics, University of British Columbia, Vancouver, British Columbia, Canada

[*] Author to whom any correspondence should be addressed: msoltani@uwaterloo.ca

Mohammad Amin Abazari's email address: m.amin.abazari@gmail.com

M. Soltani's email address: msoltani@uwaterloo.ca

Faezeh Eydi's email address: faezeheydi77@gmail.com

Arman Rahmim's email address: arman.rahmim@ubc.ca

Farshad Moradi Kashkooli's email address: farshad.moradi1987@gmail.com

# Abstract

$^{18}$F-Fluoromisonidazole ($^{18}$F-FMISO) is a highly promising positron emission tomography radiopharmaceutical for identifying hypoxic regions in solid tumors. This research employs spatiotemporal multi-scale mathematical modeling to explore how different levels of angiogenesis influence the transport of radiopharmaceuticals within tumors. In this study, two tumor geometries with heterogeneous and uniform distributions of capillary networks were employed to incorporate varying degrees of microvascular density. The synthetic image of the heterogeneous and vascularized tumor was generated by simulating the angiogenesis process. The proposed multi-scale spatiotemporal model accounts for intricate physiological and biochemical factors within the tumor microenvironment, such as the transvascular transport of the radiopharmaceutical agent, its movement into the interstitial space by diffusion and convection mechanisms, and ultimately its uptake by tumor cells. Results showed that both quantitative and semi-quantitative metrics of $^{18}$F-FMISO uptake differ spatially and temporally at different stages during tumor growth. The presence of a high microvascular density in uniformly vascularized tumor increases cellular uptake, as it allows for more efficient release and rapid distribution of radiopharmaceutical molecules. This results in enhanced uptake compared to the heterogeneous vascularized tumor. In both heterogeneous and uniform distribution of microvessels in tumors, the diffusion transport mechanism has a more pronounced than convection. The findings of this study shed light on the transport phenomena behind $^{18}$F-FMISO radiopharmaceutical distribution and its delivery in the tumor microenvironment, aiding oncologists in their routine decision-making processes.

## 1. Introduction

Solid tumor hypoxia and angiogenesis are two interdependent hallmarks in cancer biology. Hypoxia, loosely defined as cellular oxygen deficiency, increases the expression of angiogenic factors such as vascular endothelial growth factors (VEGF) and Angiopoietin-2. As such, a reciprocal relationship is induced between solid tumor hypoxia and angiogenesis in cancer progression [1]. The adverse effects of either hypoxia or uncontrolled angiogenesis can significantly alter microvascular density (MVD) in tumors leading to increased interstitial pressure, decreased drug delivery, sustained nutrient deprivation and hence tumor cell survival [2]. Therefore, the connection between solid tumor hypoxia and angiogenesis provides important implications for MVDs in clinical situations, indicating how monitoring these variables may help inform oncologic treatments [3]. In this regard, measuring hypoxia in solid tumors is essential to providing effective cancer treatment strategies that take into account both physiological and biochemical factors present in the tumor microenvironment (TME).

There are several methods available to measure and quantify the hypoxic adaptation of a species such as pO2 histograms, Oxygen extraction fraction-magnetic resonance imaging (OEF-MRI) with MRI contrast-enhancing agents, and tissue imaging by positron emission tomography (PET) with hypoxia-sensitive radiopharmaceuticals [4]. Each of these techniques relies on biomarkers or a combination of biomarkers to enable the detection, visualization, and quantification of tumor hypoxia in different organs. Among them, PET provides a more powerful tool for quantifying hypoxic tumor cells using $^{18}$F-fluoromisonidazole ($^{18}$F-FMISO).

Using this technique, it is possible to evaluate and map out the amount of oxygen present in tumors [5]. Studies assessing [18]F-FMISO PET uptake in relation to hypoxia have highlighted its utility for accurate quantification of hypoxic regions compared to other modalities [5]. Amongst its benefits, PET imaging is known to be a reliable marker of metabolic activity regardless of tissue type and gives the opportunity for longitudinal assessment in a single scan [6].

Static and dynamic imaging are two quantitative evaluation methods used to measure PET radiopharmaceutical concentrations. Static PET imaging utilizes a single time point scan of the radiopharmaceutical, while dynamic PET imaging is characterized by multiple measurements of the radiopharmaceutical concentrations over time [7-10]. Dynamic PET imaging techniques provide information about the amount of a particular radiopharmaceutical in different locations of the body at different time points. In fact, while static PET imaging provides a snapshot of metabolic activity, dynamic PET imaging, especially when used longitudinally, allows for a more comprehensive understanding of temporal changes in physiological processes, making it valuable for research and clinical applications. The dynamic PET data are then used to create and fit kinetic compartment models, which represent mathematically how the radiopharmaceutical moves and interacts between tissues, organs, and the bloodstream over time. Subsequently, they enable estimations of functional parameters such as fractional blood volume, tumor glucose utilization, and proliferation rate, as well as other pharmacokinetic parameters [11].

At the same time, kinetic compartment models commonly involve ordinary differential equations (ODEs) with differentiation with respect to time [12, 13], and do not directly model variability with respect to space within a given tumor volume. Spatiotemporal distribution models (SDMs), on the other hand, enable better quantification of tumor microenvironmental features such as perfusion, metabolism, and volume heterogeneity through partial-volume correction algorithms. These models account for the unpredictable spatial variations of radiopharmaceutical concentrations throughout different regions of tumors, including their uptake kinetics in heterogeneous tumors and metastatic regions, whereas kinetic compartment models commonly consider average values across voxels in each region. It is possible to perform kinetic modeling at the individual voxel level, referred to as parametric PET imaging [14]; however these models commonly do not explicitly model differentiation with respect to space (e.g. do not account for diffusion or convention phenomena). In addition, since the SDMs utilize partial differential equations (PDEs), beyond ODEs, which are commonly used in pharmacokinetics models, are able to calculate the radiopharmaceutical concentration over both time and space. Although SDMs have extensively been used to quantify the delivery of anti-tumor agents into solid tumors [2, 15-18], they have received less attention in the field of diagnostic agent delivery to assess PET radiopharmaceutical diffusion and uptake in solid tumors.

A few studies in the past have investigated the spatial and temporal distributions of PET radiopharmaceuticals [19-22] taking both dimensions into account. Nonetheless, the impacts of interstitial fluid fields and lymphatic drainage system were commonly not taken into account. Some used microscopic techniques involving only radial molecular diffusion [20, 21] or simplified microvessel architecture [23-26]. Additionally, these studies did not take into account convection process from vessels to tissue or within the tissue, which could affect the modeling of radiopharmaceuticals in specific areas of interest. We developed two other SDMs simulating hypoxia-PET radiopharmaceutical uptake distributions in two capillary networks [27] of a solid tumor and normal tissues around it as well as in a real-like human capillary network [28]. However, these studies did not evaluate the impact of tumors with different MVDs on semi-quantitative metrics of [18]F-FMISO. In a different context, we developed [29] a detailed SDM to analyze the uptake of the PET radiopharmaceutical [18]F-fuorodeoxyglucose ([18]F-FDG) in static and dynamic microvascular networks. The model could take into account physiological factors, such as microvessel conductivity, transvascular exchange permeability, and interstitial fluid flow fields in both healthy and cancer tissues. Later, we modified [30] this approach with a synthetic tumor microvasculature to study [18]F-FDG uptake in both healthy and tumor tissues. Moreover, some recent studies used a similar mathematical model to study the semi-quantitative [31, 32] and quantitative [33] measurements of [18]F-FDG uptake in various microvascular networks with different tumor sizes and MVDs.

To the best of our knowledge, no studies have conducted a comparison of semi-quantitative metrics for [18]F-FMISO uptake patterns in solid tumors. Additionally, there is still no research on the contribution of each transport mechanism towards [18]F-FMISO delivery to tumors with varying MVDs *via* the SDM. With this motivation, we aimed in this work to develop a novel computational framework based on intravascular injection of hypoxia PET-radiopharmaceutical to evaluate [18]F-FMISO distributions at two progression stages of solid tumors. The presented smulti-scale mathematical model, unlike many conventional compartmental models, uses PDEs to accurately calculate the distribution of [18]F-FMISO agents over both time and space. The model involves spatially correlated modeling incorporation of diffusion and convection mechanisms while considering intravascular and interstitial flows in two tumors with heterogeneous and uniform distribution of microvessels.

## 2. Methodology

### 2.1. Governing equations

#### 2.1.1. Angiogenesis model

A discrete model is being used presently to examine the connection between endothelial cells (ECs) and tumor microenvironments when it comes to tumor-induced angiogenesis. This model was pioneered by Anderson and Chaplain [34] which was then further expanded by Soltani and Chen [35] towards more realistic capillary network models. An advanced mathematical model of angiogenesis has been constructed in this study that considers characteristics such as matrix density, intravascular blood flow, anastomosis, and vessel branching. The previous works of our group contain detailed information on sprouting algorithms, equations, and governing rules [2, 35].

#### 2.1.2. Interstitial fluid flow transport

Solid tumors and the surrounding healthy tissues can be viewed as porous media [2, 17]. In a steady state condition, the momentum equation in such an environment is simplified to the Darcy law, which is expressed by [2]:

$$\vec{V}_i = -\kappa \nabla P_i \qquad (1)$$

where $\vec{V}_i$, $P_i$, and $\kappa$ are interstitial fluid velocity (IFV), interstitial fluid pressure (IFP), and hydraulic conductivity of interstitium, respectively.

The continuity equation for interstitium space in the presence of the source/sink terms caused by blood microvessels and lymphatic drainage systems is modified as follows [2]:

$$\nabla . \vec{V}_i = \underbrace{\phi_v}_{\text{Source term}} - \underbrace{\phi_l}_{\text{Sink term}} \tag{2}$$

The rate of solute transport between the blood vessels and interstitial space can be determined by calculating $\phi_v$ and $\phi_l$, which signifies the rate of solute transport per unit volume from the blood vessels to the interstitial space and from the interstitial space to lymph vessels, respectively [31].

$$\phi_v = L_P \frac{S}{V}(P_B - P_i - \sigma_s(\pi_B - \pi_i)) \tag{3}$$

$$\phi_l = L_{PL}(\frac{S}{V})_L(P_i - P_L) \tag{4}$$

Intravascular blood pressure is represented by $P_B$, the plasma osmotic pressure is denoted by $\pi_B$, the interstitial fluid osmotic pressure is represented as $\pi_i$, and the hydrostatic pressure of the lymphatic vessel is expressed as $P_L$. $\frac{S}{V}$ represents the surface area of the blood microvessels per unit volume of tissue and $L_{PL}(\frac{S}{V})_L$ indicates the loss flow rate due to the lymphatic drainage system. Table S1 provides a detailed definition of all the parameters and their corresponding values used in the simulations.

### 2.1.3. Hypoxia PET radiopharmaceutical transport

In this study, PDEs were used to model the transportation of $^{18}$F-FMISO, instead of the typical ODEs which are often included in the modeling processes for molecular imaging agents. Employing PDEs enabled us to accurately determine the spatial and temporal distribution of $^{18}$F-FMISO in both cancerous and healthy tissues. This type of mathematical model is referred to a convection-diffusion-reaction (CDR) equation, a common tool employed in describing drug delivery to solid tumors [31]. The modeling of $^{18}$F-FMISO radiopharmaceutical transport involves considering a range of biological and physiological factors, such as diffusion in microvessels, transvascular exchange between microvessels and interstitium, diffusion and convection through the interstitial spaces, and cellular uptake. The CDR equations used to represent this system are illustrated as follows [27].

$$\frac{\partial C_f}{\partial t} = \underbrace{D_{eff}\nabla^2 C_f}_{\text{Diffusion in tissue}} - \underbrace{V_i \nabla C_f}_{\text{Convection in tissue}} + \underbrace{\Phi_V}_{\text{Production rate from blood vessel}} \tag{5}$$
$$- \underbrace{\Phi_L}_{\text{Uptake rate by lymphatic vessel}} \underbrace{-K_{on}C_f + K_{off}C_b}_{\text{Exchange rate into and out of the cell}}$$

$$\frac{\partial C_b}{\partial t} = K_{on}C_f - K_{off}C_b \tag{6}$$

$$K_{on} = (\frac{K_{max}P_1}{P_{O_2} + P_1})(\frac{P_{O_2}}{P_{O_2} + P_2})^K \tag{7}$$

$$Pe = \frac{\phi_v(1 - \sigma_f)}{P_m S/V} \tag{8}$$

where $C_f$ and $C_b$ indicate the concentration of free and bound $^{18}$F-FMISO, respectively, and $D_{eff}$ stands for the effective diffusion coefficient. $K_{on}$ and $K_{off}$ denote the transport rate constants between free and bound compartments. . It is worth noting that the value of $K_{off}$ in associated studies is considered zero [27, 36, 37]. It means that once 18F-FMISO binds to the hypoxic tissue, it does not readily dissociate. This simplification is based on the assumption that 18F-FMISO, once trapped in hypoxic tissue, remains bound for a sufficiently long duration relative to the time scale of the imaging study. $P_{O_2}$ represents the oxygen pressure distribution in tissue, which is considered to be 40 mmHg [27]. Moreover, according to previous mathematical [27] and experimental [38] studies, since the oxygen pressure decreased to about 0 mmHg at a distance of 150 µm from the microvessel wall, it has been assumed in the radiopharmaceutical distribution modeling that the oxygen pressure in tissues is equal to the microvascular oxygen pressure [27]. $P_1$ is the oxygen pressure in tissue that inhibits binding by 50% necrosis, while $P_2$ is the oxygen pressure that causes 50% necrosis. $K$ indicates step width at $P_2$ and $K_{max}$ represents the maximum binding rate. $Pe$ is the Peclet number which demonstrates the significance of convective transport relative to diffusive transport. $P_m$ stands for the transvascular permeability of the vessel to the radiopharmaceutical and $\sigma_f$ is the filtration reflection coefficient.

The transport rate of a radiopharmaceutical from microvessels into the interstitial space is indicated by $\Phi_V$, while the transport rate of the radiopharmaceutical from the interstitial space into lymphatic microvessels is represented by $\Phi_L$. According to Patlak's model, these two terms can be calculated as [27]:

$$\Phi_V = L_P \frac{S}{V} \underbrace{\left( P_B - P_i - \sigma_s (\pi_B - \pi_i) \right)}_{\text{Convection from vessel wall}} \left( 1 - \sigma_f \right) C_p \qquad (9)$$

$$+ \underbrace{\left( C_p - C_f \right)}_{\text{Diffusion from vessel wall}} \frac{P_m \frac{S}{V} Pe}{\exp(Pe) - 1}$$

$$\Phi_L = \phi_l C_f \qquad (10)$$

where $C_p$ is the plasma intravascular concentration of 18F-FMISO agents. The value and definition of 18F-FMISO radiopharmaceutical transport parameters are listed in Table S2.

### 2.1.4. Semi-quantitative assessment of 18F-FMISO uptake

The standardized uptake value (SUV) index, which is ubiquitous in oncological practices, is calculated for semi-quantitative analysis of changes in radiopharmaceutical uptake during tumor progression. The SUV is determined by the total tissue radioactivity concentration to the injected radioactivity, normalized by the body weight like the case of a 75 kg patient [31]. Additionally, $C_t$ (total 18F-FMISO concentration) is calculated as the sum of both concentration terms ($C_f$ and $C_b$) [31].

$$\text{SUV} = \frac{\text{Total concentration } (C_t)}{\text{Injected radioactivity}} \times \text{Body weight} \qquad (11)$$

$$C_t = C_f + C_b \qquad (12)$$

### 2.2. Computational domain and boundary conditions

The schematic of the computational domain in the presence of a vascularized tumor is shown in Figure 1. Here, a two-dimensional computational domain is considered at which the square domain indicates the healthy tissue with a length of L = 5 cm and a solid tumor with a diameter of $D_{\text{Tumor}}$ = 2 cm is placed at $(\frac{L}{2}, \frac{3L}{4})$. Here, to evaluate TME factors affecting hypoxia PET radiopharmaceutical delivery and its uptake, two tumor networks are used. One vascularized tumor network with heterogeneous distribution of capillary network (i.e., tumor I) is generated by solving systems of equations related to the tumor-induced angiogenesis process. The other tumor network is a uniformly distributed vascularized tumor (i.e., tumor II), in which its capillary network is uniformly spread throughout the domain to demonstrate the extreme effect of MVD. In tumor I, the microvascular network grows toward the tumor tissue from a single parent vessel located on the lower vertical line of the domain, where five sprouts initiated the angiogenesis process.

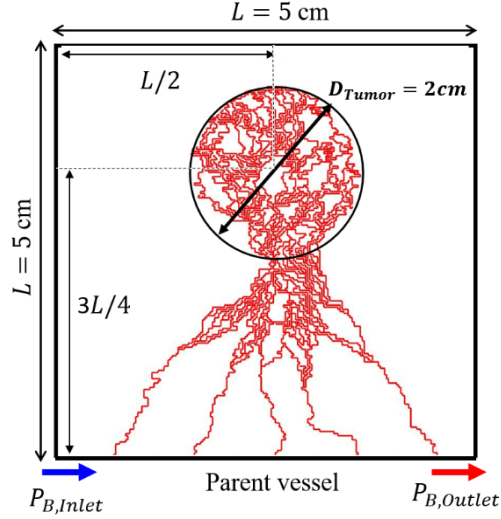

Figure 1. Illustration of the computational domain specification used in the present study.

For simulating intravascular blood flow, the inlet and outlet pressure boundary conditions are required. These values have been chosen according to physiological-accurate boundaries outlined in the literature [27], as follows:

$$\text{Parent vessel} \rightarrow \left\{ \begin{array}{c} P_{B,Inlet} = 25 \ (\text{mmHg}) \\ P_{B,Outlet} = 5 \ (\text{mmHg}) \end{array} \right\}$$

At the interface between tumor and healthy tissue (i.e., the inner boundary), continuity boundary conditions for interstitial fluid flow and $^{18}$F-FMISO concentration are chosen [31] as:

$$(-\kappa_t P_i |_{\Omega^t}) = (-\kappa_n P_i |_{\Omega^n}) \qquad (13)$$
$$(P_i |_{\Omega^t}) = (P_i |_{\Omega^n})$$
$$(-D_{eff}^t \nabla C + v_i C) \lfloor \Omega^t = (-D_{eff}^n \nabla C + v_i C) \lfloor \Omega^n \qquad (14)$$

$$(C |_{\Omega^t}) = (C |_{\Omega^n})$$

where $C$ indicates the $^{18}$F-FMISO concentration and $\Omega^n$ and $\Omega^t$ represent the healthy and tumor tissues at their interface, respectively.

The IFP at the outer boundary has a constant value; therefore, a Dirichlet boundary condition is imposed on the edges of the domain [2, 31] as:

$$P_i = \text{Constant} \qquad (15)$$

The open boundary condition is applied to the domain edges to facilitate mass transport across boundaries, whereby convective inflow and outflow can take place [31]. The condition is represented by Eq. (16) and is defined by the normal vector $n$.

$$-n.\nabla C = 0 \qquad (16)$$

## 2.3. Solution strategy and computational simulation

The methodology used in this study is outlined in Figure 2. To start, a lattice-discrete probabilistic model [2, 35] is utilized to generate tumor-associated vasculature with adjustable microvessels and non-continuous blood flow. Afterward, the Hagen-Poiseuille equation is used to solve the fluid flow in microvessels and to find the blood pressure distributions through capillary networks. Subsequently, the mass and momentum equations in the interstitium are numerically solved using obtained intravascular blood pressure distributions from the previous step. Then, IFP and IFV values are implemented to solve CDR equations. Finally, [18]F-FMISO radiopharmaceutical concentrations are considered for obtaining SUV-dependent parameters for both stages of tumor growth.

The computational fluid dynamics software COMSOL Multiphysics 5.6 (COMSOL Inc., Burlington MA, USA) was employed to solve the governing equations relevant to interstitial fluid flow and transportation of [18]F-FMISO, such as the continuity, Darcy and CDR equations; these are solved using finite element methods with four error magnitudes corresponding to residual squares. Darcy's Law is applied to solve the continuity and Darcy equations, while the General Form PDE (physics) can be utilized to calculate the CDR equations with Lagrangian shape functions and quadratic components used for each of the PDEs. Furthermore, COMSOL Multiphysics software 5.6 also has an 'Interpolation Function' which allows users to import blood pressure fields into the program. For computation purposes, a direct solver labeled as MUMPS - multifrontal massively parallel sparse direct solver - works hand in hand with the backward differentiation formula time; stepping process having a time step size of 1 second. Decreasing the step size has no significant effects on outcomes however it does influence the amount of time taken for complete computation.

To compare the results of IFP and intracellular [18]F-FMISO radiopharmaceutical concentration obtained with four distinct computational grids (coarse, medium, fine, and extremely fine), a grid independent test is conducted. Variations lower than 5% between medium and fine grids, as well as lower than 2% among corresponding fine and extremely fine grids, are noted representing a benefit from the lowest computational costs exerted by finer grids. Therefore, the fine grid is considered followed by subsequent simulations using about 26,000 triangular elements including 320 edged elements for an average element quality of 0.90. All the simulations are done utilizing a personal laptop that has an Intel (R) Core i7-8550U processor, 1.80 GHz CPU, and 16 GB of RAM.

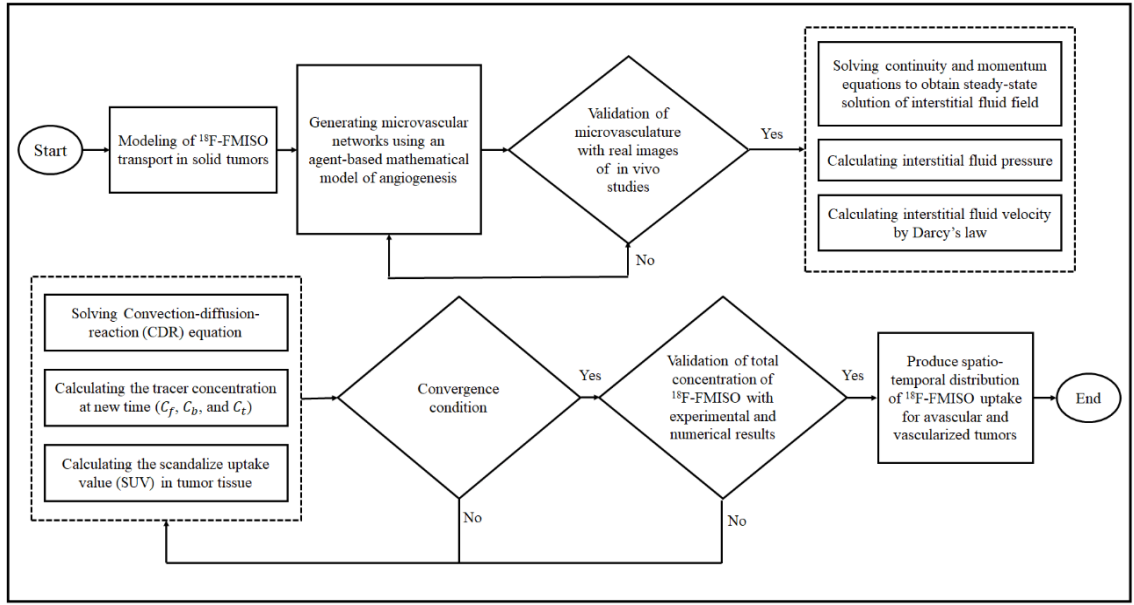

Figure 2. Step-by-step representation of the used algorithm for spatiotemporal modeling of the $^{18}$F-FMISO radiopharmaceutical.

## 2.4. Validation of the model

The current computational model has been validated by comparison with numerical results and experimental data from previously published studies. The microvascular network is compared with biological data of solid tumor morphology qualitatively. Our mathematical modeling has successfully generated microvascular networks that match with the in vivo observations [39-41]. These in vivo observations show that new vessels tend to branch out more near the periphery of the tumor and within the tumor domain. This is due to the high gradient of tumor angiogenetic factors within the tumor domain, which is relatively higher than the surrounding healthy tissue. In addition, the growth of microvessels from initial capillary sprouts is validated qualitatively by the in vivo tumor-induced capillary architectures.

Figure 3 shows the comparison of the mean total concentration of $^{18}$F-FMISO agents in the vascularized tumor with other experimental analyses as well as numerical results. Bruehlmeier et al. [42] performed dynamic $^{18}$F-FMISO PET scans in a patient with glioblastoma to measure tumor hypoxia and perfusion, while Soltani et al. [42] used an SDM to explore the time activity curves of $^{18}$F-FMISO agents in the vascularized solid tumors. Overall, there is a consistent trend between our results and experimental data as well as average concentration found numerically, providing evidence for the robustness of our mathematical model in considering the complexity of both biological and physiological factors of the TME. There is a slight difference between our data and the experimental one in the first 20 minutes, which is due to differences in boundary conditions, computational domain, tumor diameters, and capillary networks. Our analysis shows that 20 minutes post-injection, the total uptake is successfully matched with both experimental and numerical data.

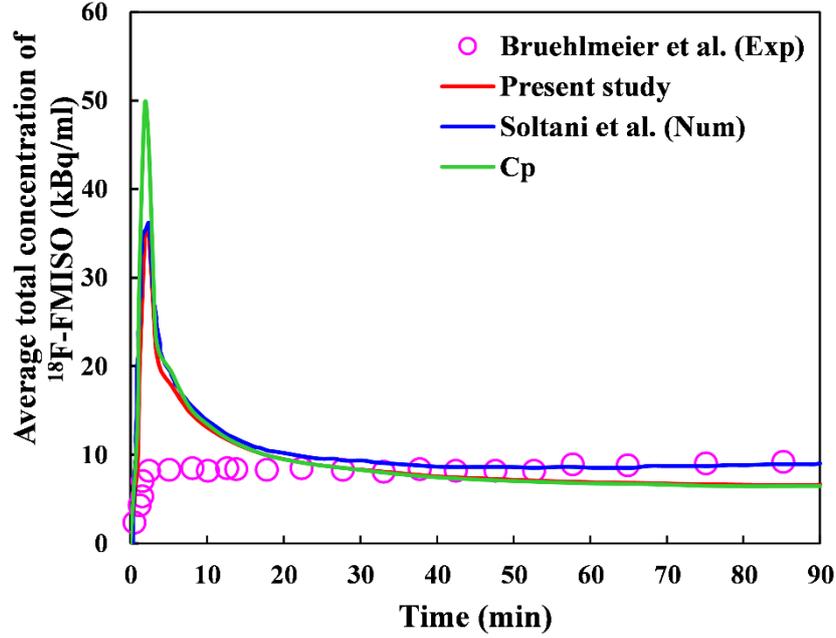

Figure 3. Comparison of the average total concentration of $^{18}$F-FMISO in the vascularized tumor of the present study against previously published numerical [42] and experimental [42] investigations. The plasma intravascular concentration of $^{18}$F-FMISO [42] ($C_p$) is also shown.

# 3. Results and discussion

In the current investigation, concentrations of $^{18}$F-FMISO agents over space and time are calculated by solving the momentum, continuity, and CDR equations in the interstitium, taking the biological factors of TME into account. Furthermore, to explore the influences of MVD on the distribution of $^{18}$F-FMISO agent and its uptake, two tumor networks have been considered. Subsequently, this research will go into greater detail by assessing the distributions of IFP and IFV, time activity curves of $^{18}$F-FMISO agents, and related SUV-based factors in both tumor and healthy tissues. Finally, the contribution of each mechanism of $^{18}$F-FMISO agents transport in tissue and microvessels is presented in the two aforementioned tumor networks.

## 3.1. Interstitial fluid flow quantities

According to Figure 4, the IFP value is at its highest in the tumor compared to healthy tissue across both tumor networks. The IFV has a very low magnitude (in the order of $10^{-8}$ ms$^{-1}$) in most parts of the domain, with the exception of a thin layer at the boundary between tumor and healthy tissue, where IFP gradients are substantial and create a maximum value. As MVD increases, the IFP in through tumor tissue also climbs, ranging from 2.04 kPa (vascularized tumor) to 2.43 kPa (uniformly vascularized tumor). In the vascularized tumor, some regions with high MVD have also a higher IFP than other areas, while the upper regions of the tumor with purely and normalized distributed microvessels have the lowest value of intratumoral IFP. The lowering of IFP due to a reduction in MVD has been noted in experimental investigations on both human tumors [43] and transplanted tumors in mice [44, 45] as well as numerical studies [2, 46]. Moreover, remarkable qualitative and qualitative agreements are found between the trend of these interstitial fluid flow quantities and their values obtained by our advanced model and the corresponding parameters obtained in the experimental data [47, 48] and the numerical investigations [2, 31, 33].

Elevated IFP in tumor tissues has an important role in the delivery of therapeutic and diagnostic agents. Elevated IFP can reduce the effectiveness of treatments that rely on the transport of solutes through the extracellular space, such as chemotherapy [2] and gene therapy. Additionally, IFP can serve as a physical obstacle to the delivery of diagnostic agents, such as MRI contrast dyes or radiation therapies [49]. From a biological point of view, the main reasons for such elevated IFP in the tumor rely on the high hyperpermeability of tumor blood vessels, contraction and stiffening of the extracellular matrix, as well as, the lack of a functional lymphatic system at the tumor tissue [2, 31].

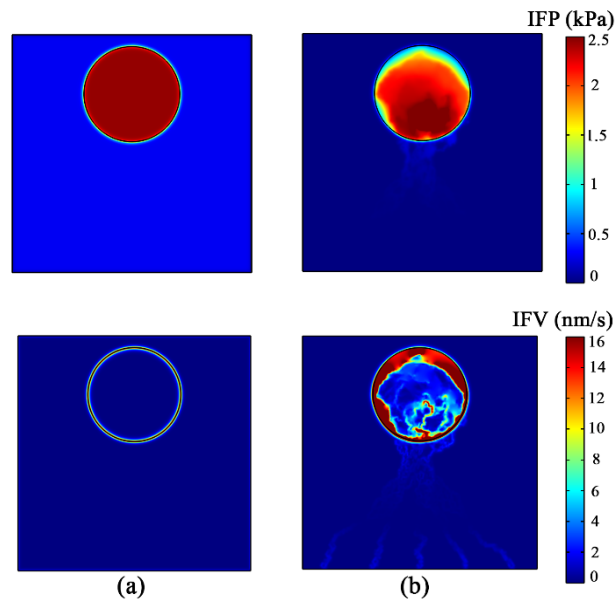

Figure 4. Illustration of interstitial fluid flow quantities. IFV and IFP distributions for (a) uniform and (b) heterogeneous distributions of microvessels in vascularized tumors predicted by the proposed model.

## 3.2. Quantitative and semi-quantitative assessments of hypoxia PET radiopharmaceutical

[18]F-FMISO PET radioactivity curves of tumor tissues are plotted against time from 0 to 90 minutes after injection in Figure 5. Overall, our computational results show that hypoxic PET radiopharmaceutical uptake highly depends on the capillary network density. The uniformly vascularized tumor predicts a higher value of intracellular and extracellular [18]F-FMISO concentrations than the vascularized tumor network. As shown in Figure 5(a), during the first two minutes of injection, extracellular [18]F-FMISO concentration shows an upward trend until reaching its maximum value. The extracellular concentration suddenly drops, and then as time goes it smoothly keeps reducing until minute 80 and then remains almost unchanged. The downward trend in the extracellular concentration explains [18]F-FMISO transporting into the bound state ($C_b$) to tumor cells with low oxygen levels within the intracellular space. Consequently, as shown in Figure 5(b), the intracellular concentration gradually increases due to the low diffusion coefficient of [18]F-FMISO which requires more time to transport within the tissue. Figure 5(c) illustrates the mean standardized uptake value (SUV$_{mean}$) which closely follows the extracellular concentration trend. Our results also demonstrate that, during the first 20 minutes of injection, the SUV$_{mean}$ has different values in tumors with different degrees of angiogenesis.

Our quantitative and semi-quantitative analyses indicate that MVD plays a crucial role in the distribution and uptake of hypoxia PET radiopharmaceutical in solid tumors. The high microvascular density presented in the uniformly vascularized tumor provides a pathway to increase [18]F-FMISO uptake, as it enables more [18]F-FMISO molecules to release and distribute into different regions of the tumor tissue in a shorter period of time. This ultimately leads to increased uptake of [18]F-FMISO both intracellularly and extracellularly compared with the vascularized tumor. Such findings are consistent with previous computational and experimental studies that investigated the transport of therapeutic and diagnostic agents into solid tumors. For instance, some recent studies mathematically demonstrated that the spatiotemporal distribution of [18]F-FDG uptake is also significantly impacted by the architecture of tumor vasculatures [29, 31, 33]; in the initial stage of tumor growth, [18]F-FDG uptake by cancerous cells is limited due to the lower microvascular density [31]. Similarly, Abazari et al. [2] showed that a greater MVD within the tumor tissues facilitates more efficient transport and distribution of anti-cancer drugs during both immunochemotherapy as well as classical chemotherapy. Chauhan et al. [50] found that normalizing blood vessels in mammary tumors with anti-angiogenic therapies enhances the delivery and effectiveness of small nanoparticles.

Moreover, a comparison between $SUV_{mean}$ values in the uniformly vascularized and vascularized tumors also indicates that the $SUV_{mean}$ of the uniformly vascularized tumor is slightly higher due to a higher level of MVD. Therefore, for accurate estimation of SUV, the $SUV_{mean}$ formula should be modified to include the microvascular density parameter, leading to a more meaningful metric for oncologists in their decision-making processes prior to routine clinical reporting. Some experimental and mathematical studies addressed the effect of tumor size/type [51] and microvascular density [31] on both the SUV and [18]F-FDG uptake.

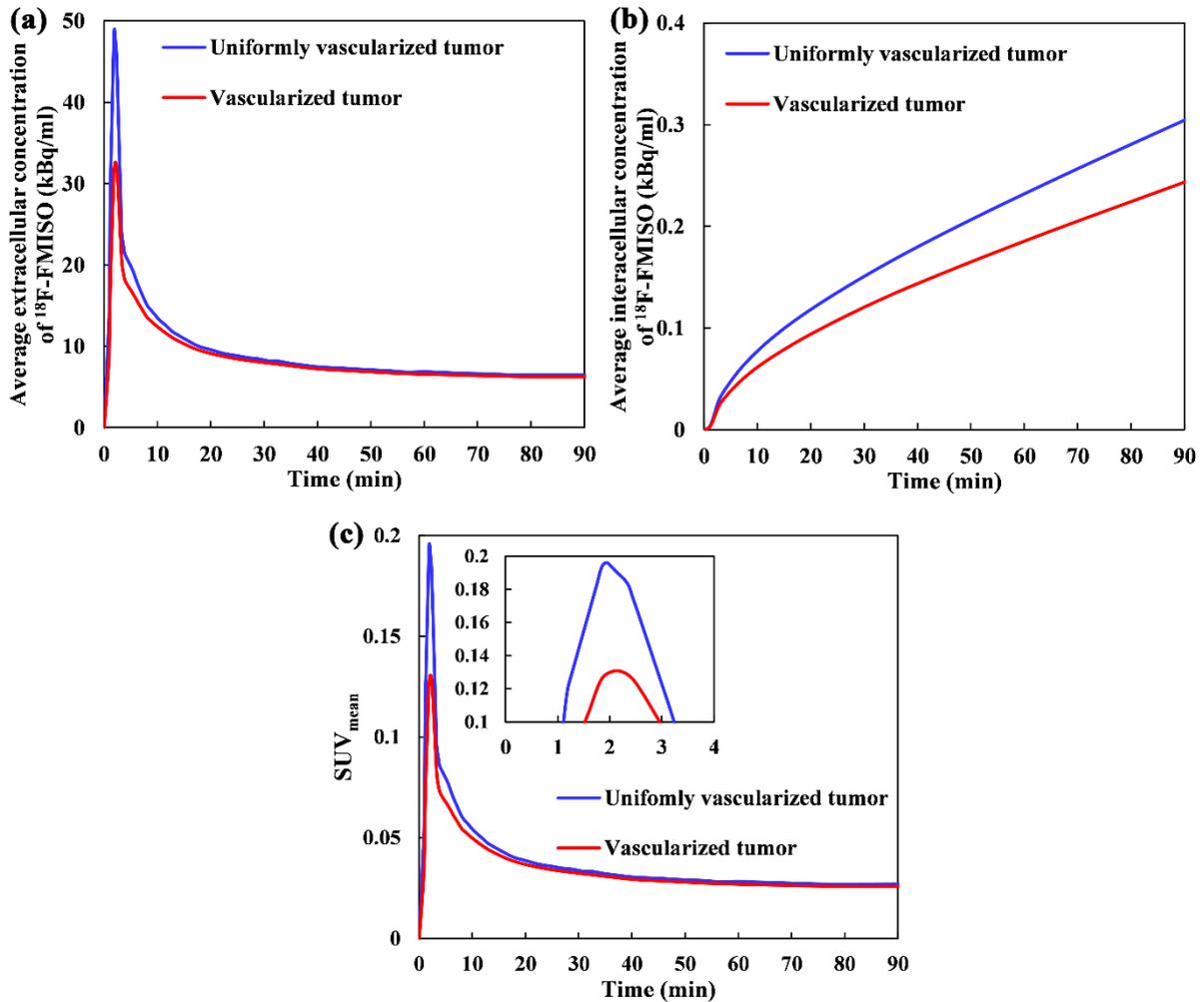

Figure 5. Time activity curves of $^{18}$F-FMISO within uniformly vascularized and vascularized tumors. (a) Extracellular $^{18}$F-FMISO concentration, (b) intracellular $^{18}$F-FMISO concentration, and (c) mean standardized uptake value (SUV$_{mean}$).

Figure 6 illustrates the extracellular, intracellular, and standardized uptake value distributions of $^{18}$F-FMISO in the vascularized tumor network at various times post-injection, using spatiotemporal distribution modeling. At first time steps, extracellular $^{18}$F-FMISO is more abundant than intracellular, however, this ratio changes as time progresses due to the molecules being converted from free to bound. High SUV and extracellular concentrations of $^{18}$F-FMISO are seen around capillaries outside of tumor tissue, indicating a high level of MVD. This can be attributed to the fact that the vessels in microvascular networks are used as sources for $^{18}$F-FMISO release, causing higher concentrations close to them. Furthermore, convection and diffusion mechanisms spread the molecules throughout other regions of the tumor domain [52]. In addition to that, as mentioned in the previous sections, the IFP is high in the center of tumors and weaker at the periphery and surrounding healthy tissue. In consequence, one would expect that the interstitial fluid moves from the edges of the tumor into the surrounding tissue. For a macromolecule to diffuse into the tumor, it needs to go beyond this outward convection [53]. The relative significance of the convection and diffusion mechanism of nonuniform distribution of $^{18}$F-FMISO in tumors may be smaller than heterogeneous extravasation as a result of elevated pressure and necrosis.

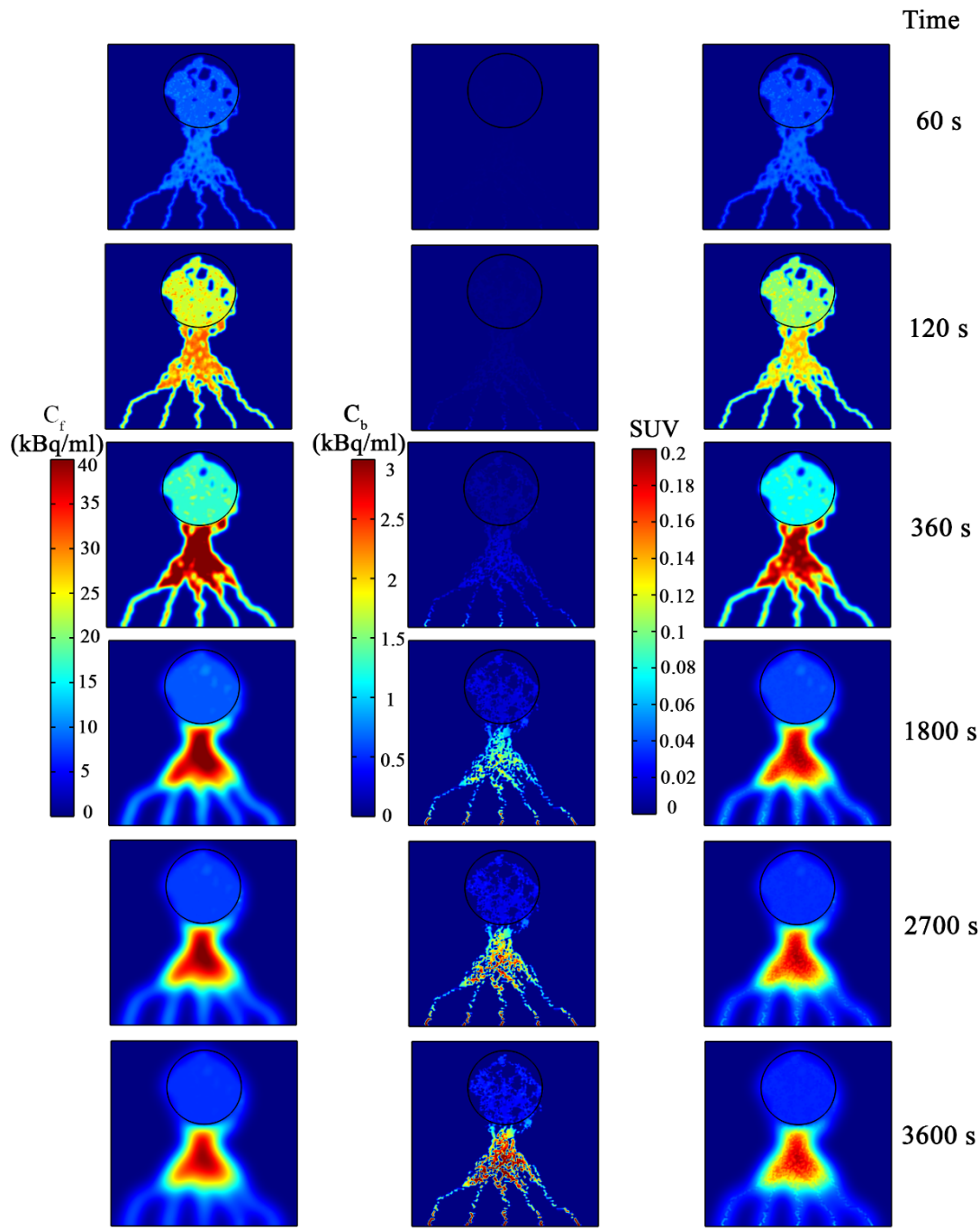

Figure 6. Series of snapshots from spatiotemporal distribution modeling of $^{18}$F-FMISO for the vascularized tumor network at various times post-injection. Columns 1-3 show the extracellular $^{18}$F-FMISO ($C_f$), intracellular $^{18}$F-FMISO ($C_b$), and SUV distributions estimated by the model, respectively. The black circular contour line indicates the boundary of the solid tumor.

### 3.3. Importance of different mechanisms on hypoxia PET radiopharmaceutical transport

Four key transport mechanisms used in our system of equations are: (1) diffusion from vessel walls, (2) convection from vessels, (3) diffusion in tissue, and (4) convection in tissue. Here, to evaluate the impact of each of these four terms on the $^{18}$F-FMISO distributions, one term is removed while the others are kept. In our CDR equations, terms related to $^{18}$F-FMISO transport from vessels to tissue and tissue to vessels, as well as radiopharmaceutical binding within the cells/matrix, are all outlined. The diffusion-based transport mechanisms of [18]F-FMISO occur due to the concentration gradient between high and low areas of PET activity. As the [18]F-FMISO molecules move from an area of higher concentration to one of lower concentration, they become evenly dispersed throughout the healthy and tumor region. On the other hand, the convection-based transport mechanisms of [18]F-FMISO are driven by pressure gradients created between areas of high and low pressure. In PET imaging, this occurs when [18]F-FMISO molecules move away from a region with higher concentration due to the push of higher pressure produced within the tumor gradients.

Based on the calculations, the contribution of four transport mechanisms on the [18]F-FMISO concentration in two tumor cases is shown in Figure 7. A comparison between the two pie charts shows that hypoxia PET radiopharmaceutical distributions highly depend on the vessel network morphology. In both tumor cases, the transport of radiopharmaceutical by diffusion is dominant over the convection-based transport terms. In the vascularized tumor, the major mechanisms are diffusion from vessels with 72% portion and diffusion in tissue with 28% portion, while the convection-based transport mechanisms contribute to lower than 1%. In contrast, in the uniformly vascularized tumor, diffusion from vessels is the only transport mechanism with 100% contribution.

In line with previous mathematical modelings [29, 37], our results indicate that convection terms have little impact on [18]F-FMISO transport, thus they can be removed from the calculation process. Omitting convection from the SDM simplifies the calculations and ultimately decreases the computational cost compared to simulations with convection consideration. In more detail, convection terms are driven by the velocity of interstitial flow as well as transvascular pressure difference. In solid tumors, such differences are negligible when compared to the diffusion-driven flow of interstitial agents. Furthermore, convection terms within solid tumors are obstructed due to solid constituents obstructing the movement of interstitial agents and cells which reduces convective forces.

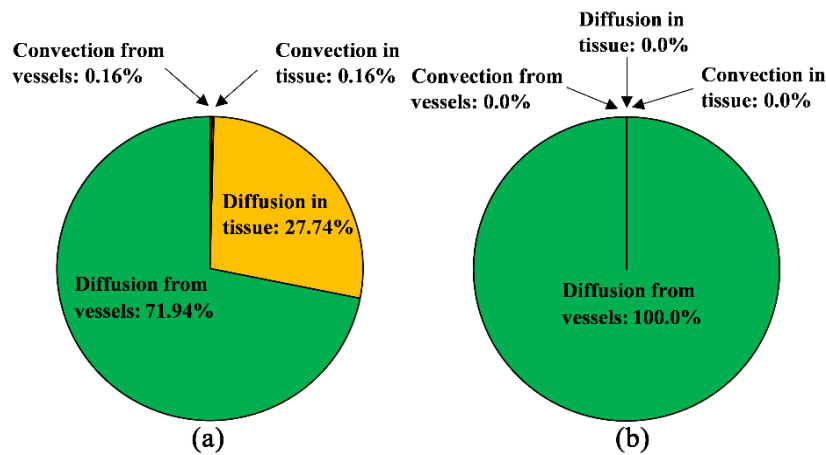

Figure 7. The contribution of different terms of CDR equations on [18]F-FMISO distribution for (a) a vascularized tumor and (b) a uniformly vascularized tumor.

On the other hand, both diffusion terms have a major effect on [18]F-FMISO transport. The magnitude of the diffusion from vessel term is based on the capillary network structure. Our findings suggest that in the dense network (uniformly vascularized tumor) with higher MVDs, the contribution of diffusion from vessel to tissue is greater than in tissue itself; however, when the density decreased, the contribution of diffusion in tissue increased. Therefore, since the capillary network structure within real tumor tissue can vary from heterogeneous or uniform, diffusion-based terms unlike convection-based terms are unable to be eliminated.

The current study has certain limitations along with assumptions that should be considered. First of all, a 2D geometry of tumor with 1D geometry of microvessels (i.e., 2D-1D model) have been investigated, while to have a comprehensive overview on the results, a full 3D geometry of tumor and microvessels is needed. However, this kind of 2D-1D medels has already shown acceptable levels of reliability compared to experimental results [2, 27, 29, 31, 32, 35]. This model is also limited by the number of tumor microvessel networks with various MVDs. The geometry of tumors can be complex due to various parameters such as irregular shape and growth pattern, infiltrative invasion into surrounding tissues, and the presence of necrotic or cystic areas within the tumor mass. However, our model is intended to contribute to the understanding of the distribution of [18]F-FMISO radiopharmaceutical agents in generic tumor tissues, independent of each individual, through convection, diffusion, and reaction mechanisms. Although the model performed well with two specific tumor geometries and their corresponding MVDs, more comprehensive investigations based on a larger cohort of tumor networks would confirm the significance of tumors with different degrees of angiogenesis in the prediction of [18]F-FMISO radiopharmaceutical distribution. In addition, while the model's ability to predict interstitial fluid parameters and time activity curves is comprehensively validated by computational and experimental studies, the model's predictions of [18]F-FMISO radiopharmaceutical uptake were not validated due to the lack of appropriate experimental data in the literature. Further analyses should expand upon this work with a combined *in silico*, *in vitr*o, *in vivo*, approach to authenticate the results' accuracy.

## 4. Conclusions

We developed a comprehensive biologically-based mathematical model to predict the spatiotemporal distribution of hypoxia PET radiopharmaceutical and its uptake in solid tumors with different degrees of angiogenesis. The comprehensive model, unlike common kinetic compartment models, determines the influence of intravascular and interstitial fluids on the transport of [18]F-FMISO radiopharmaceutical within the tumor and its surrounding healthy tissues. It also considers the underlying TME factors, such as microvessel conductivity and transvascular exchange permeability. Its ability to calculate interstitial fluid parameters as well as time activity curve of the [18]F-FMISO radiopharmaceutical is validated by qualitative and quantitative comparisons with experimental and numerical data. Results show that tumor II has a larger MVD through which more [18]F-FMISO agents can extravasate, penetrate and bind to the tumor cells, resulting in higher cellular uptake values than those observed in tumor I. A comparison between $SUV_{mean}$ values in two stages of tumor shows that the $SUV_{mean}$ of tumor II is slightly higher due to a higher level of MVD. This implies the need to revise the SUV formula to include the MVD parameter to obtain an accurate estimation of SUV, thus providing a more meaningful metric for decision-making processes by oncologists. The results of the analysis on the impact of different mechanisms on radiopharmaceutical transport imply that [18]F-FMISO distributions are highly dependent on the capillary network architecture. In both heterogeneous and uniformly distributed microvessels, the transport of radiopharmaceutical by diffusion transport mechanisms is dominant over the convection-based terms. As such, in spatiotemporal modeling of hypoxia PET radiopharmaceutical, convection-based terms unlike diffusion-based terms can be excluded to simplify the calculations and ultimately decrease the computational cost. The proposed mathematical model will serve as an intuitive approach to explore the impact of tumor MVD in the spatiotemporal modeling in PET imaging of hypoxia.

**Abbreviations**

VEGF: vascular endothelial growth factors

MVD: microvascular density

TME: tumor microenvironment

OEF-MRI: Oxygen extraction fraction-magnetic resonance imaging

PET: positron emission tomography

$^{18}$F-FMISO:$^{18}$F-fluoromisonidazole

ODEs: ordinary differential equations

SDMs: spatiotemporal distribution models

PDEs: partial differential equations

$^{18}$F-FDG: $^{18}$F-fuorodeoxyglucose

ECs: endothelial cells

IFV: interstitial fluid velocity

IFP: interstitial fluid pressure

SUV: standardized uptake value

CDR: convection-diffusion-reaction

**Declarations**

**Ethics approval and consent to participate**

Not applicable.

**Consent for publication**

Not applicable.

**Availability of data and material**

The datasets used and/or analysed during the current study are available from the corresponding author on reasonable request.

**Competing interests**

The authors declare that they have no competing interests.

**Funding**

Not applicable.

**Acknowledgements**

Not applicable.